\documentclass[
aps,
prl,
twocolumn,
amsmath,
amssymb,
graphicx,
showpacs,
floatfix,
bibnotes,
]
{revtex4-1} 

\newcommand{\bcen}{\begin{center}}
\newcommand{\ecen}{\end{center}}
\newcommand{\btab}{\begin{tabular}}
\newcommand{\etab}{\end{tabular}}
\newcommand{\bdes}{\begin{description}}
\newcommand{\edes}{\end{description}}

\newcommand{\beq}{\begin{equation}}
\newcommand{\eeq}{\end{equation}}
\newcommand{\bea}{\begin{eqnarray}}
\newcommand{\eea}{\end{eqnarray}}

\newcommand{\bary}{\begin{array}}
\newcommand{\eary}{\end{array}}
\newcommand{\benum}{\begin{enumerate}}
\newcommand{\eenum}{\end{enumerate}}
\newcommand{\bitem}{\begin{itemize}}
\newcommand{\eitem}{\end{itemize}}

\newcommand{\br} { {\boldsymbol{r}}}

\newcommand{\bzero} { {\boldsymbol{0}}}

\newcommand{\D}[1]{\mbox{d}{#1}}

\newcommand{\ket}[1]{{| #1 \rangle}}

\newcommand{\Fig}[1]{Fig.~\ref{#1}}

\newcommand{\AII}{{\sf{AII}}}

\newcommand{\C}{{\sf{C}}}

\newcommand{\DIII}{{\sf{DIII}}}

\newcommand{\mylabel}[1]{\label{#1}}

\usepackage{hyperref}
\usepackage{amsmath}
\usepackage{epsfig}
\usepackage{wrapfig} 
\usepackage{mathrsfs}
\usepackage{amssymb} 
\usepackage{color}
\usepackage{amsbsy}
\usepackage{subfigure}
\usepackage{marginnote}
\usepackage{slashed}
\usepackage{marvosym}
\usepackage{pifont}
\usepackage{mathbbol}
\usepackage{wrapfig}
\usepackage{upgreek}

\newcommand{\usenomenclature}{}
\ifdefined\usenomenclature
\usepackage[noprefix]{nomencl}
\fi

\let\chapter\section
\let\section\subsection
\let\subsection\subsubsection

\newcommand{\oibook}[1]{}

\renewcommand{\le}{\leqslant}

\newcommand{\mytitle}{Topological Insulators in Random Lattices}

\newcommand{\myaffl}{{Department of Physics, Indian Institute of Science, Bangalore 560012, India.}}

\begin{document}

\title{\mytitle}
\author{Adhip Agarwala}
\author{Vijay B.~Shenoy}\email{shenoy@physics.iisc.ernet.in}
\affiliation{\myaffl}

\date{\today{}}
\begin{abstract} 
Our understanding of topological insulators is based on an underlying crystalline lattice where the local electronic degrees of 
freedom at different sites hybridize with each other in ways that produce nontrivial band topology, and the search for 
material systems to realize such phases have been strongly influenced by this. Here we theoretically demonstrate topological insulators in systems with a random distribution of sites in space, i.~e., a random lattice.  This is achieved by constructing hopping models on random lattices whose ground states possess nontrivial topological nature (characterized e.~g., by Bott indices) that manifests  as quantized conductances in systems with a boundary. By tuning parameters such as the density of sites (for a given range of fermion hopping), we can achieve transitions from trivial to topological phases. We discuss interesting features of these transitions. In two spatial dimensions, we show this for all five symmetry classes (A, AII, D, DIII and C) that are known to host nontrivial topology in crystalline systems.   We expect similar physics to be realizable in any dimension and provide an explicit example of a $Z_2$ topological insulator on a random lattice in three spatial dimensions. Our study not only provides a deeper understanding of the topological phases of non-interacting fermions, but also suggests new directions in the pursuit of the laboratory realization of topological quantum matter. 
\end{abstract}


\maketitle 

\noindent
{\bf Introduction:} The band insulating state of many fermions has received renewed attention in recent times\cite{Hasan_RMP_2010,Qi_RMP_2011, Ando_JPSJ_2013}. Clues that the band insulating state may support additional nontrivial physics -- attributed to topology -- was provided by the discovery of the integer quantum Hall effect\cite{Klitzing_PRL_1980} and the theoretical work that followed\cite{Laughlin_PRB_1981,Thouless_PRL_1982,Haldane_PRL_1988}. These ideas saw a resurgence with the discovery of the two dimensional spin Hall insulator\cite{Murakami_PRL_2004,Kane_PRL_2005a,Kane_PRL_2005b,Bernevig_PRL_2006,Bernevig_Science_2006,Konig_Science_2007a}, soon followed\cite{Fu_PRL_2007,Moore_PRB_2007,Roy_PRB_2009b,Hsieh_Nature_2008} by the three dimensional topological insulator (see \cite{Hasan_RMP_2010,Qi_RMP_2011, Ando_JPSJ_2013}).

A complete classification of gapped phases of non-interacting fermions soon appeared\cite{Qi_PRB_2008,Schnyder_PRB_2008,Ryu_NJP_2010,Kitaev_AIP_2009}. This classification hinges on the ten symmetry classes of Altland and Zirnbauer\cite{Altland_PRB_1997}. In any given spatial dimension $d$ only five symmetry classes of the ten host topologically nontrivial phases (see e.g~Kitaev\cite{Kitaev_AIP_2009}). Two gapped systems are considered to be topologically equivalent if the ground state of one can be reached starting from the ground state of the other by a symmetry preserving adiabatic deformation of the Hamiltonian which does not close the gap during the deformation process. The ground state of a gapped crystalline system is characterized by a set of filled bands. For each point in the Brillouin zone (BZ), this amounts to a Slater determinant state made of Bloch wavefunctions (whose character is determined by the symmetry class) of the filled bands. In $d$-dimensions this turns out to be a map from the BZ ($d$-dimensional torus) to points on a symmetric space whose character is determined by the total number of bands and the symmetry of the system \cite{Kitaev_AIP_2009}. Whether such a map allows for nontrivial ``winding'' decides if a symmetry class supports topological phases in that dimension. Well known lattice models of Haldane \cite{Haldane_PRL_1988},  Kitaev \cite{Kitaev_PU_2001}, Kane-Mele \cite{Kane_PRL_2005a}, BHZ \cite{Bernevig_Science_2006}, etc. fall into this paradigm. Gapped topological phases realized in such systems are robust to impurities and disorder (that preserve the symmetry), and their surfaces typically host gapless modes which open possible new directions for technological applications\cite{Hasan_RMP_2010,Qi_RMP_2011, Ando_JPSJ_2013}. These ideas have strongly influenced the experimental search of topological systems both in material\cite{Das_NatPhys_2012, Chang_Science_2013, Nadj_Science_2014} and synthetic cold-atomic systems \cite{Jotzu_Nature_2014}.

While robustness to disorder is the defining property of any topological state, the caveat which continues to be applied to all studies is ``disorder should not be large enough to close the gap". Most studies which have attempted to address the question of disorder in topological systems have, therefore, an implicit crystalline lattice and a ``small" disorder. The effect of onsite(Anderson) disorder on a $d=3$ topological insulator was studied in ref.~\cite{Kobayashi_PRL_2013}, where it was shown that with increasing disorder a topological state transits to a metal. The effect of disorder in Kitaev chains has also been studied\cite{Diez_NJP_2014}. Other pertinent studies include, how an onsite disorder can induce\cite{Li_PRL_2009} a topological phase in a trivial system (which has the necessary ingredients to produce topological phases), and an interesting concept of ``statistical topological insulators'' which requires another statistical symmetry apart from the symmetry protecting the topological phase\cite{Fulga_PRB_2014}. Questions related to the robustness against disorder have also been posed in the context of weak topological insulators where it has been shown that they are surprisingly robust to disorder than previously expected\cite{Ringel_PRB_2012}.
Physics of topological phases in noncrystalline lattices have received relatively less attention. Most studies have focused on quasicrystalline systems\cite{Kraus_PRL_2012, Fulga_PRL_2016,Bandres_PRX_2016} , for example, realizing a weak topological insulator phase in such a system\cite{Fulga_PRL_2016}.
An interesting unexplored question, both from theoretical and practical perspectives, is whether a completely random set of points i.~e., a random lattice, such as that realized by impurities in a material, can host topological phases. This is the question that we address in this paper. Random lattices have previously been explored in the context of the fermion doubling problem\cite{Christ_NPB_1982} of lattice field theory, however, the possibility of topological phases in them has not been so far addressed.

In this paper, we theoretically establish that random lattices can host topologically insulating phases. We provide a demonstration of this by constructing models (using familiar ingredients) on random lattices where fermions hop to sites within a finite range. By tuning parameters (such as the density of sites), we show that the system undergoes a quantum phase transition from a trivial to a topological phase. We characterize the topological nature by obtaining the topological invariant (cf. Bott index\cite{Loring_EPL_2010}), and associated quantized transport signatures. We also address interesting features of such quantum phase transitions. This is achieved through a detailed study of all nontrivial symmetry classes (A, AII, D, DIII and C) in two dimensions. We also provide a demonstration of a $Z_2$ topological insulator in three dimensions. We believe that this work opens a new direction in the experimental search for topological quantum matter, for example in impurity band of a three dimensional band insulator (realization in $d=3$), and randomly placed atoms on a two dimensional insulating surface ($d=2$ realization), and even amorphous or ``glassy" materials.



\begin{figure}
\includegraphics[width=8cm]{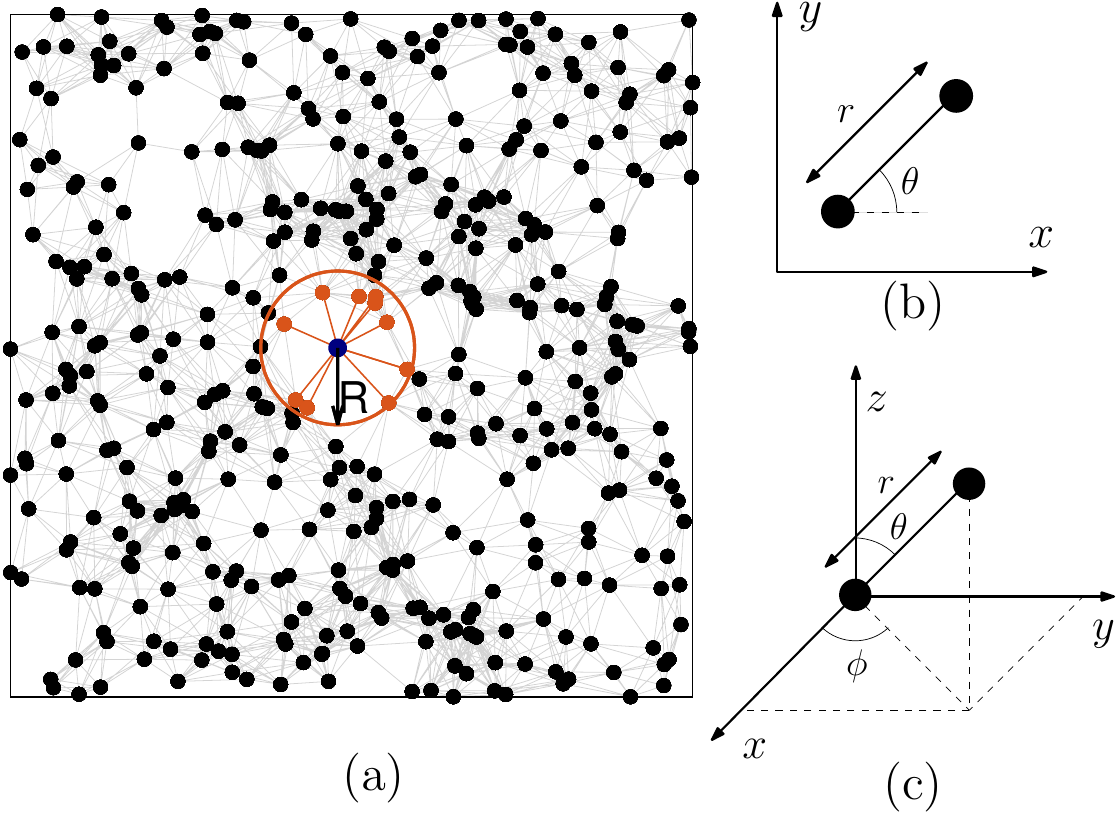}
\caption{{\bf Random lattice:} (a) A configuration of random lattice. Sites are indicated by dark spots. Light lines indicate the hoppings between the sites which are within the distance $\le R$ from each other. (b,c) The separation between two sites is described by a distance $r$ and an angle $\theta$ in two dimensions (b) and by $(\theta, \phi)$ in three dimensions (c).}
\label{fig:TypLat}
\end{figure}

\noindent
{\bf Model:} We consider a region(box) of $d$-dimensional space whose volume is $V$. Our random lattice is constructed by placing $N$ sites (labeled by $I=1,\ldots,N$) randomly in the box (see \Fig{fig:TypLat}). The positions of the sites are sampled from an uncorrelated uniform distribution. The collection of points is characterized by a single parameter, viz.,
\beq\mylabel{eqn:rho}
\rho = \frac{N}{V}.
\eeq

The sites are identical, each of which hosts $L$ single particle states. The states at the site $I$ are denoted by $\ket{I \alpha}$. The associated fermion operators are denoted by $c^\dagger_{I,\alpha}$ and $c_{I,\alpha}$ ($\alpha=1,\ldots,L$). The ``flavor'' label $\alpha$ can stand for the spin quantum number or an orbital quantum number as appropriate.

Fermions can hop from site to site, possibly changing their flavor in the process. We mimic a realistic system by considering a finite range $R$ of hopping (see \Fig{fig:TypLat}). A generic Hamiltonian of such a system is
\beq\mylabel{eqn:HamGen}
{\cal H} = \sum_{I\alpha} \sum_{J \beta} t_{\alpha\beta}(\br_{IJ})c^\dagger_{I,\alpha}c_{J, \beta}. 
\eeq
Here, $J$ runs over all sites that are within a distance of $R$ of $I$, i.~e., $|\br_{IJ}| \le R$ (with $\br_{IJ}$ being the vector from $I$ to $ J$). The $L \times L$ matrix $t_{\alpha \beta}(\br)$ depends on the vector $\br =r \hat{\br}$ via
\beq\mylabel{eqn:tab}
t_{\alpha \beta}(\br) = t(r) T_{\alpha \beta}(\hat{\br})
\eeq

The distance dependence of the hopping is captured by $t(r)$ and $T_{\alpha\beta}(\hat{\br})$ contains both the orbital and angular dependencies. This form is motivated by hoppings found in spin-orbit coupled systems that are common in the realization of topological phases. The unit vector $\hat{\br}$ in two spatial dimensions is described by an angle $\theta$ with respect to a chosen axis (see \Fig{fig:TypLat}(b)), while in three dimensions we parametrize $\hat{\br}$ by polar and azimuthal angles $(\theta,\phi)$. The hopping matrix for $\br = \bzero$ 
\beq\mylabel{eqn:onsite}
t_{\alpha \beta}(\bzero) = \epsilon_{\alpha \beta}
\eeq
describes the ``onsite energy'' or the ``atomic'' Hamiltonian of the system. For, $r \ne 0$, we choose
\beq\mylabel{eqn:tr}
t(r) = C \Theta(R - r) e^{-r/a} 
\eeq
where the constant $C$ is chosen such that $t(r)$ is a unit energy when $r=a$, i.~e., $C=e$. The step function $\Theta$ enforces the cutoff distance $R$. In the construction that follows, the forms of $\epsilon_{\alpha \beta}$ and $T_{\alpha \beta}(\hat{\br})$ will be motivated by systems that are experimentally relevant\cite{Bernevig_Science_2006, Qi_RMP_2011}. Finally, to investigate the physics of topological edge states we will study system with and without periodic boundary conditions on the box. In remaining discussion the scale $a$ is set to unity, and all other lengths are measured in units of $a$.

\begin{widetext}
\begin{center}
\begin{table}
\begin{tabular}{|c|c|c|}
\hline
\shortstack{\\Class\\(par)} & $\epsilon_{\alpha \beta} $ & $T_{\alpha \beta}(\hat{\br})$  \\ 
\hline 
\shortstack{\\A\\$(\lambda,M,t_2)$}
&
$\begin{pmatrix}
 2+M  &    (1-i) \lambda  \\
(1+i) \lambda & -(2+M) 
\end{pmatrix}$  
& 
$\begin{pmatrix}
\frac{-1+t_2}{2}  & \frac{-ie^{-i\theta} +\lambda(\sin^2 \theta(1+i)-1)}{2} \\
\frac{-ie^{i\theta}+\lambda (\sin^2 \theta(1-i)-1)}{2} &   \frac{1+t_2}{2}
\end{pmatrix}$ \\
\hline
\shortstack{\\AII\\$(\lambda, M$,\\$ t_2,g)$ }
&
$\begin{pmatrix}
 2+M+2t_2  &-i2\lambda & 0 & 0 \\
i 2 \lambda & -(2+M)+2t_2 & 0 & 0 \\
 0 & 0 & 2+M+2t_2 & i2\lambda  \\
0 & 0 & -i 2 \lambda & -(2+M) + 2t_2
\end{pmatrix}$
&
$ \begin{pmatrix}
-\frac{1}{2} - \frac{t_2}{2}  & -\frac{i}{2}e^{-i\theta} + \frac{i\lambda}{2} & 0 & -\frac{ig}{2}e^{-i\theta}\\
-\frac{i}{2}e^{i\theta} - \frac{i\lambda}{2} &   \frac{1}{2} - \frac{t_2}{2} & -\frac{ig}{2}e^{-i\theta} & 0\\
0 &  -\frac{ig}{2}e^{i\theta} & -\frac{1}{2} - \frac{t_2}{2}  & \frac{i}{2}e^{i\theta} - \frac{i\lambda}{2} \\
-\frac{ig}{2}e^{i\theta} & 0 & \frac{i}{2}e^{-i\theta} + \frac{i\lambda}{2}&   \frac{1}{2} - \frac{t_2}{2}
\end{pmatrix}
$ \\ 
\hline 
\shortstack{\\D\\$(\mu, \Delta)$}&
$\begin{pmatrix}
 2-\mu  &    0  \\
0 & -(2- \mu) 
\end{pmatrix}$
&
$\begin{pmatrix}
 -\frac{1}{2}  &  \Delta e^{i\theta} \\
 -\Delta e^{-i\theta}&   \frac{1}{2}
\end{pmatrix}$ \\ 
\hline 
\shortstack{\\DIII \\ $(M,g)$} &
AII$(\lambda=0, t_2=0)$ & AII$(\lambda=0, t_2=0)$  \\ 
\hline 
\shortstack{\\C\\$(M)$}
& 
$\begin{pmatrix}
 2+M  &    0  \\
0 & -(2+M) 
\end{pmatrix}$
& 
$\begin{pmatrix}
 -\frac{1}{2} & -\frac{1}{2}e^{-i 2\theta}  \\
 -\frac{1}{2}e^{i 2\theta} & \frac{1}{2}
\end{pmatrix} $ \\ 
\hline 
\end{tabular}
\caption{{\bf Hamiltonians:} Onsite energies and hopping matrix elements in different symmetry classes in two dimensions. The size of the matrix determines the value of $L$ in each class. The Cartan labels of each class and the set of independent parameters(par) in the Hamiltonian are shown in column 1. The model of class A is inspired by  anomalous quantum Hall effect \cite{Bernevig_Book_2013}, AII by $Z_2$ topological insulator model \citep{Bernevig_Science_2006}, D by $p+ip$ superconductor \cite{Bernevig_Book_2013}, DIII by time-reversal invariant superconductor \cite{Roy_arXiv_2006,Qi_PRL_2009} and C by $d+id$ superconductor \cite{Senthil_PRB_1999,Chern_AIP_2016}.}
\label{2DRanHam}
\end{table}
\end{center}
\end{widetext}

\noindent
{\bf Two dimensional systems:} We begin the discussion of our results with two dimensional systems. As is known there are five symmetry classes  (A, AII, D, DIII and C) that allow  topological phases to exist in two dimensions. We construct Hamiltonians for each of these classes that respect all the relevant symmetries, as shown in Table.~\ref{2DRanHam}. We have performed analysis of systems in all these five classes, and discuss the results of class A in detail as a representative.

The symmetry class A possesses no intrinsic symmetries like time reversal, charge conjugation, or sublattice. We obtain a Hamiltonian in class A with two orbitals on every site and characterized by three parameters $\lambda, t_2$ and $M$ as shown in Table.\ref{2DRanHam}. In fact, this structure can be obtained by starting from the BHZ model \cite{Bernevig_Science_2006,Bernevig_Book_2013} by introducing $t_2$ which breaks charge conjugation symmetry. Note also that $\lambda$ breaks the spatial inversion symmetry. $M$ (mass parameter) is the parameter we tune to investigate the possibility of topological phases for various values of $\rho$, the density of sites.

\begin{figure}
\includegraphics[width=8cm]{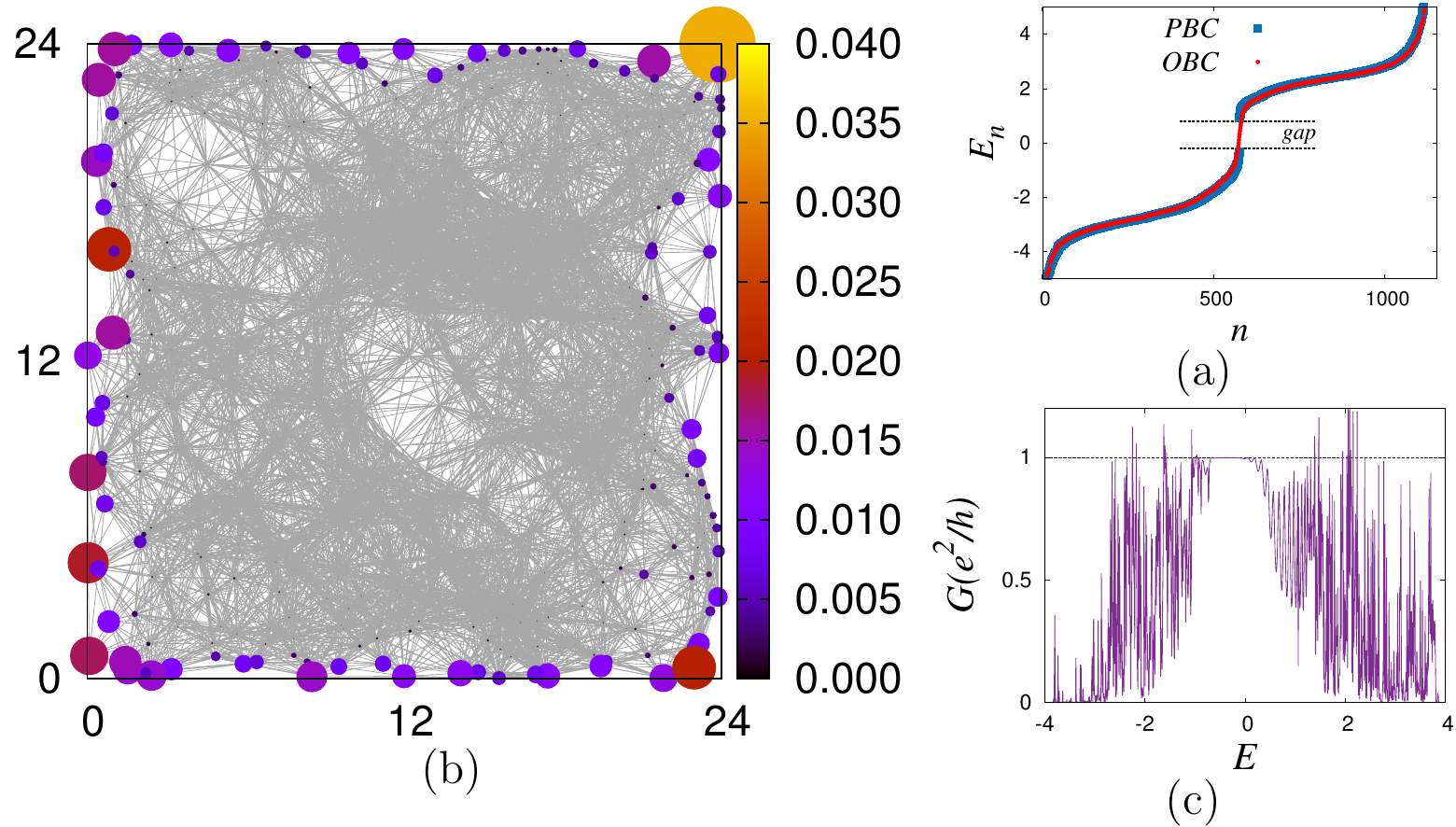}
\caption{{\bf Class A random lattice model ($d$=2):} (Area $24\times 24, R=4, M=-0.5,t_2=0.25,\lambda=0.5,\rho=1$)  (a) Energy eigenvalues $E_n$ versus the state number $n$. The system with periodic boundary conditions(PBC) shows a gap; while that with open boundaries(OBC) shows mid-gap states. (b) The wavefunction of the mid-gap state localized on the edge. The size and the color of the blob indicates the probability of finding a fermion at that site. (c) Two terminal conductance ($G$) as a function of incident fermion energy showing a quantized value in the energy gap. }
\label{fig:ClassArandm}
\end{figure}

Results of a particular realization of the random lattice for a $24\times24$ system with $\rho=1$, $R=4$  and $M=-0.5$ is shown in \Fig{fig:ClassArandm}. \Fig{fig:ClassArandm}(a) shows the energy eigenvalues of the system with and without the periodic boundary conditions. In the presence of periodic boundary condition the system clearly shows an energy gap i.~e, it is an insulator. In the absence of periodic boundary condition (in both directions) we see a set of energy eigenvalues in the mid-gap region.  A typical mid-gap state, we find, is an ``edge" state which is localized on the ``surfaces" of the box(see \Fig{fig:ClassArandm}(b)). We studied the transport due to this edge state by coupling the system to leads along opposite surfaces while the other two were open, i.~e, they possess edge states. \Fig{fig:ClassArandm}(c) shows the two terminal conductance that we calculate using the non-equilibrium Green's function formalism \cite{Datta_Book_1997}. Remarkably, when the energy of the incoming fermions is in the ``bulk gap" we see the conductance is quantized to unity.

\begin{figure}
\includegraphics[width=8cm]{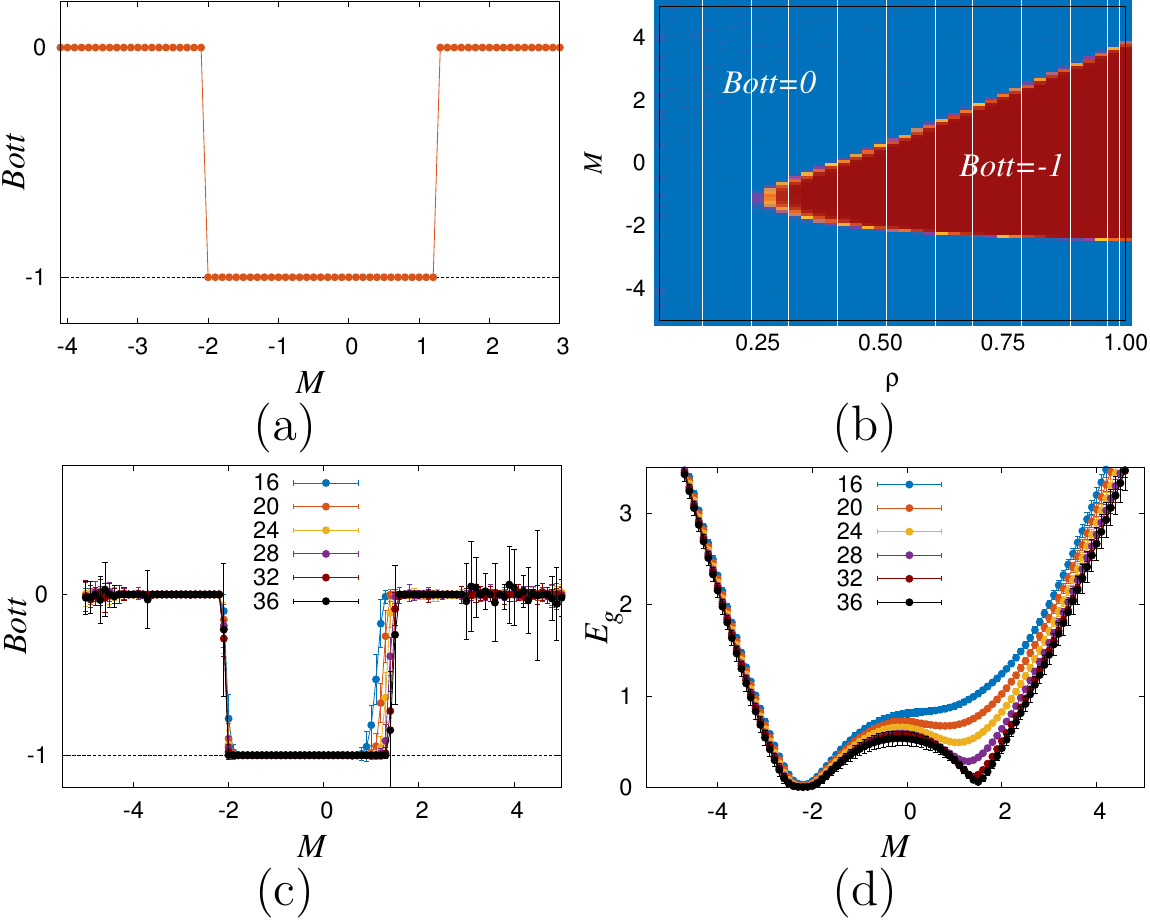}
\caption{{\bf Phase diagram:}(a) Bott index for a particular realization of the random lattice at $\rho=0.6$. (Area $24\times 24, R=4,t_2=0.25,\lambda=0.5$)  (b) Contour plot of (configuration averaged) Bott index in the $M-\rho$ plane. Red region indicates topologically non-trivial phase. (c) Configuration averaged Bott index for various system sizes. (d) Configuration averaged energy gap $E_g$ for various system sizes. (c) and (d) are also for $\rho=0.6$. Configuration average is performed over 320 realizations of the random lattice. Other parameters are kept same as in (a).}
\label{fig:ClassABott}
\end{figure}

\begin{figure}
\includegraphics[width=8cm]{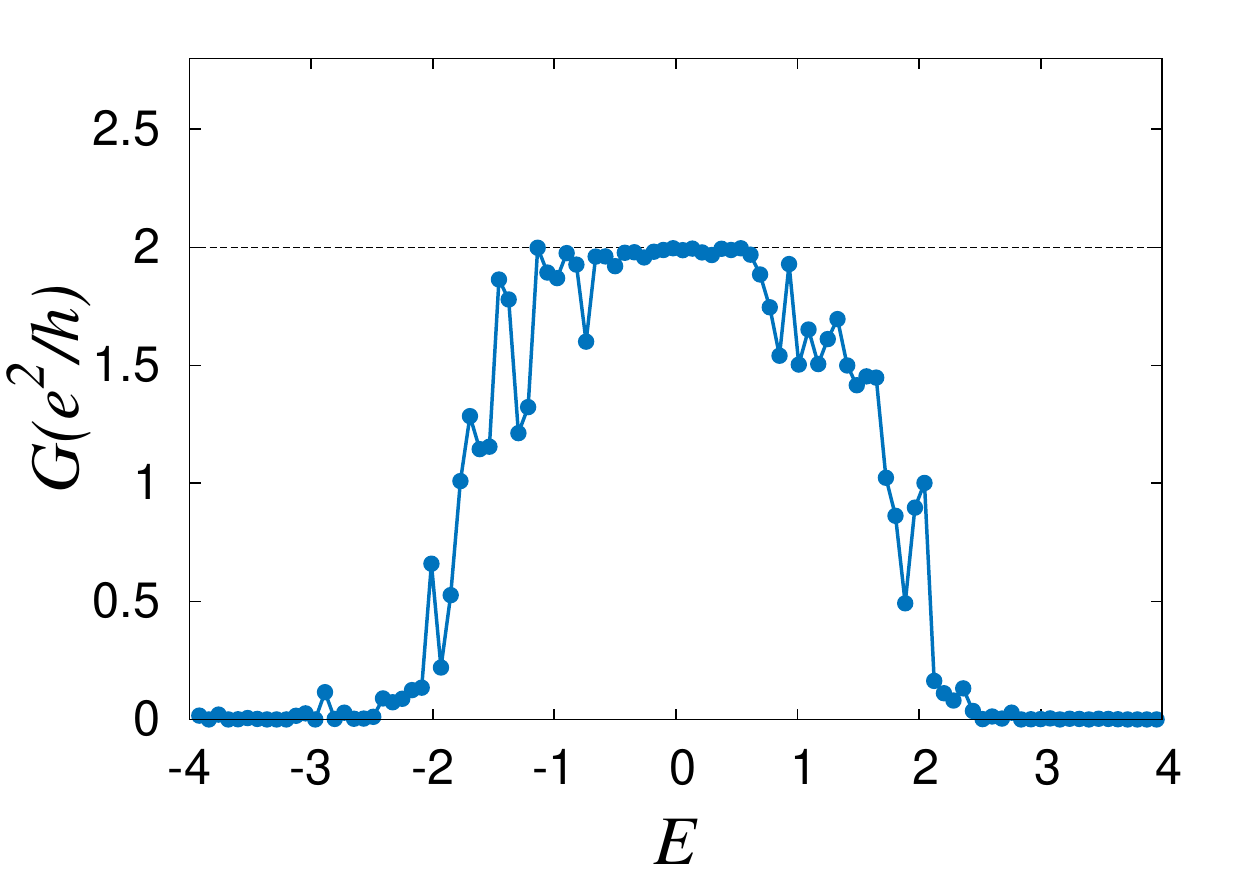}
\caption{{\bf Class C:} Topological phase in class C on the random lattice. The two terminal conductance is shown as a function of incident fermion energy for a single random lattice configuration($M=-1,V=60\times 60, \rho=1, R=4$). The plateau at $2e^2/h$ can be clearly seen. We find the Bott index to be $=-2$.}
\label{fig:ClassC}
\end{figure}

Are the edge states that we see of topological origin? The presence of an edge state that provides for a quantized conductance in a completely random lattice, if surprising, is suggestive of its topological origin. To confirm this we calculate a topological invariant, the Bott index, following Loring and Hastings \cite{Loring_EPL_2010}. Remarkably the system shows a Bott index of negative unity confirming the topological character of the ground state. Under what condition does a random lattice show topological phases? We address this question by obtaining a phase diagram in the $M-\rho$ plane, i.~e, by varying the mass parameter and the density of sites. \Fig{fig:ClassABott}(a) shows the Bott index for a particular configuration as a function of the mass parameter $M$. For $-2 \lesssim M \lesssim 1.2$ the system is in a topological phase with two quantum phase transitions at $M \approx -2$ and at $M \approx 1.2$. To obtain the phase diagram we average over 320 configurations (so the values shown are averaged over 320 configurations at each value of density). \Fig{fig:ClassABott}(b) shows a contour plot the Bott index over the $M-\rho$ plane. We see that there is a large regime of parameters in the density and $M$ where the system is topological. An important point is that a critical density $\rho_c$ is needed to obtain such a topological phase. The existence of such a critical density is expected to be ``universal", although the value will be determined by the specific microscopic parameters.
 For any $\rho>\rho_c$, note that there are two values of $M$ at which the system has a phase transition. To investigate the nature of these phase transitions we studied the scaling properties of the gap and the Bott index as a function of the system size. Upon reducing $M$ from a positive value, in the first transition that is encountered, the gap does not vanish in a system of a finite size, and indeed the smallest gap as a function of $M$ reduces with increasing system size. This indicates that the states involved in this transition are ``long wavelength" modes i.~e., states whose wavefunctions are extended over the sample. Indeed this is also indicated in the jump of the Bott index which becomes increasingly sharper with increasing system size. This story is quite different at the second transition that occurs at a positive $M$. Here the gap vanishes and does not scale with the system size. Similarly the sharpness in jump of the Bott index does not scale with the system size. 
 As mentioned similar physics occurs in other symmetry classes. We show in \Fig{fig:ClassC} results for class C whose Bott index is an even number and has an even quantized conductance.

\begin{figure}
\includegraphics[width=8cm]{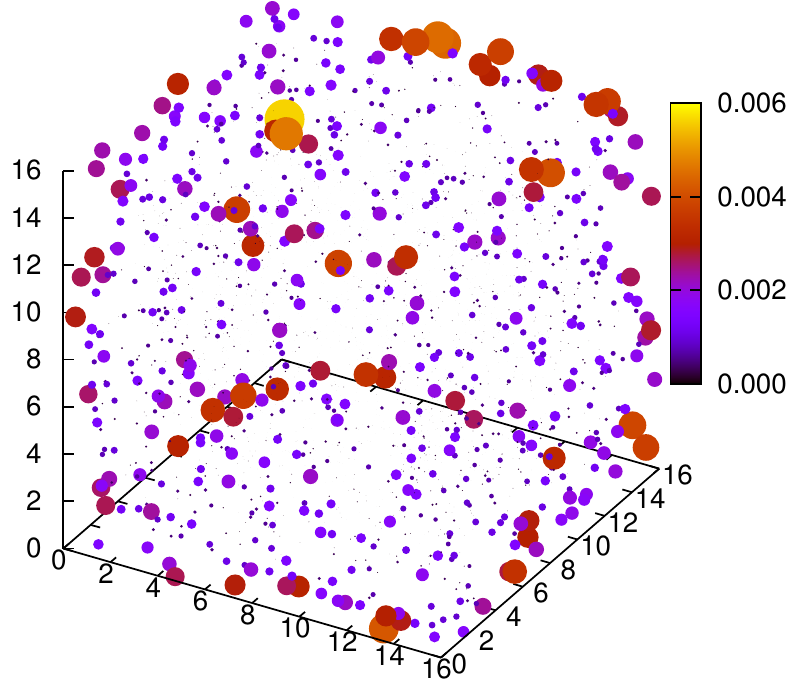}
\caption{{\bf ${\bf Z_2}$ system in three dimensions:} The mid-gap state localized on the surface. The size and the color of the blob indicate the probability of finding a fermion at the site. 
($\rho=0.6, V=16^3, M=0.5, R=4$)}
\label{fig:3Drandm}
\end{figure}

\noindent
{\bf Three dimensional system:} Given the vast interest enjoyed by the three dimensional topological insulator (a system with a $Z_2$ invariant), we also investigate the possibility of realizing this in a random lattice. To this end, we consider a system with four orbitals ($L=4$) with a Hamiltonian described by $\epsilon_{\alpha \beta}=\mbox{Diag}(-3+M,-3+M,3-M,3-M)$ where $M$ is the mass parameter, and
{\footnotesize
\beq
T_{\alpha \beta}(\theta,\phi)= 
\frac{1}{2}
\begin{pmatrix}
1 & 0 & -i\cos \theta & -ie^{-i\phi}\sin \theta\\
0 & 1 & -ie^{i\phi}\sin \theta & i\cos \theta\\
-i \cos \theta & -i e^{-i\phi}\sin \theta & -1  & 0 \\
-ie^{i\phi}\sin \theta & i\cos \theta & 0 &  -1
\end{pmatrix}
\eeq
}
This system has the required time reversal symmetry. For an appropriate set of parameters ($M=0.5$ and $\rho=0.6$), we find indeed that there are mid-gap states in an open system whose wavefunctions are localized on the boundary (see \Fig{fig:3Drandm}). This clearly indicates the new possibilities for creating topological phases in three dimensions.

\noindent
{\bf Perspective:} The possibility of topological phases in a completely random system opens up several avenues both from experimental and theoretical perspectives. Our results suggest some new routes to the laboratory realization of topological phases. First, two dimensional systems can be made by choosing an insulating surface on which suitable ``motifs'' (atoms/molecules) with appropriate orbitals are deposited at random (note that this process will require far less control than conventional layered materials). The electronic states of these motifs will then hybridize to produce the required topological phase. Second is the possibility of creating three dimensional systems starting from a suitable large band gap trivial insulator. The idea then is to place ``impurity atoms'', again with suitable orbitals and ``friendly'' chemistry with the host, not unlike the process  of $\delta$-doping of phosphorus in silicon\cite{Scappucci_APL_2009}. The hybridization of the impurity orbitals would again produce a topological insulating state in the impurity bands under favorable conditions. Here we do not thoroughly explore the specific material system that will realize these ideas. In realistic systems the temperature scales over which one will see the topological physics (determined by the bandwidth and band gap) may be low. Nevertheless, we believe that our results will motivate materials science experts to address these challenges. Finally, our work also suggests that a completely amorphous system can be a topological insulator, clearly providing new opportunities to search for ``glassy'' insulator with spin-orbit coupled motifs to realize topological phases.

Our work also provides some interesting new directions for theoretical research. There are two (equivalent) ways to view topological phases. The first one is ``kinematic'', i.~e., based on the homotopy of ground state wavefunctions of systems in a given symmetry class as discussed in the introduction. The second approach, probably better suited for the random system, is based on asking if the $d-1$ dimensional surface of a gapped $d$-dimensional system resists localization\cite{Ryu_NJP_2010}. This absence of localization on the $d-1$ surface can be used to characterize the topology of the $d$ dimensional bulk. Localization is prevented by the presence of a topological term in the action (nonlinear $\sigma$ model) that describes the low energy modes of the $d-1$ dimensional surface. While one usually writes down such $\sigma$-models based on symmetry considerations, an interesting question in the current context would be to uncover how such topological terms can arise in the random lattice setting.


\noindent
{\bf Acknowledgments:} AA thanks CSIR, India, for support and VBS acknowledges a research grant from SERB, DST, India. The authors thank Arijit Haldar and Aabhaas Mallik for comments on the manuscript.

\bibliography{RandomTI}

\begin{thebibliography}{44}%
\makeatletter
\providecommand \@ifxundefined [1]{%
 \@ifx{#1\undefined}
}%
\providecommand \@ifnum [1]{%
 \ifnum #1\expandafter \@firstoftwo
 \else \expandafter \@secondoftwo
 \fi
}%
\providecommand \@ifx [1]{%
 \ifx #1\expandafter \@firstoftwo
 \else \expandafter \@secondoftwo
 \fi
}%
\providecommand \natexlab [1]{#1}%
\providecommand \enquote  [1]{``#1''}%
\providecommand \bibnamefont  [1]{#1}%
\providecommand \bibfnamefont [1]{#1}%
\providecommand \citenamefont [1]{#1}%
\providecommand \href@noop [0]{\@secondoftwo}%
\providecommand \href [0]{\begingroup \@sanitize@url \@href}%
\providecommand \@href[1]{\@@startlink{#1}\@@href}%
\providecommand \@@href[1]{\endgroup#1\@@endlink}%
\providecommand \@sanitize@url [0]{\catcode `\\12\catcode `\$12\catcode
  `\&12\catcode `\#12\catcode `\^12\catcode `\_12\catcode `\%12\relax}%
\providecommand \@@startlink[1]{}%
\providecommand \@@endlink[0]{}%
\providecommand \url  [0]{\begingroup\@sanitize@url \@url }%
\providecommand \@url [1]{\endgroup\@href {#1}{\urlprefix }}%
\providecommand \urlprefix  [0]{URL }%
\providecommand \Eprint [0]{\href }%
\@ifxundefined \urlstyle {%
  \providecommand \doi  [0]{\begingroup \@sanitize@url \@doi}%
  \providecommand \@doi [1]{\endgroup \@@startlink {\doibase
  #1}doi:\discretionary {}{}{}#1\@@endlink }%
}{%
  \providecommand \doi  [0]{doi:\discretionary{}{}{}\begingroup
  \urlstyle{rm}\Url }%
}%
\providecommand \doibase [0]{http://dx.doi.org/}%
\providecommand \Doi [0]{\begingroup \@sanitize@url \@Doi }%
\providecommand \@Doi  [1]{\endgroup\@@startlink{\doibase#1}\@@Doi}%
\providecommand \@@Doi [1]{#1\@@endlink}%
\providecommand \selectlanguage [0]{\@gobble}%
\providecommand \bibinfo  [0]{\@secondoftwo}%
\providecommand \bibfield  [0]{\@secondoftwo}%
\providecommand \translation [1]{[#1]}%
\providecommand \BibitemOpen [0]{}%
\providecommand \bibitemStop [0]{}%
\providecommand \bibitemNoStop [0]{.\EOS\space}%
\providecommand \EOS [0]{\spacefactor3000\relax}%
\providecommand \BibitemShut  [1]{\csname bibitem#1\endcsname}%
\bibitem [{\citenamefont {Hasan}\ and\ \citenamefont
  {Kane}(2010)}]{Hasan_RMP_2010}%
  \BibitemOpen
  \bibfield  {author} {\bibinfo {author} {\bibfnamefont {M.~Z.}\ \bibnamefont
  {Hasan}}\ and\ \bibinfo {author} {\bibfnamefont {C.~L.}\ \bibnamefont
  {Kane}},\ }\Doi {10.1103/RevModPhys.82.3045} {\bibfield  {journal} {\bibinfo
  {journal} {Rev. Mod. Phys.},\ }\textbf {\bibinfo {volume} {82}},\ \bibinfo
  {pages} {3045} (\bibinfo {year} {2010})}\BibitemShut {NoStop}%
\bibitem [{\citenamefont {Qi}\ and\ \citenamefont {Zhang}(2011)}]{Qi_RMP_2011}%
  \BibitemOpen
  \bibfield  {author} {\bibinfo {author} {\bibfnamefont {X.-L.}\ \bibnamefont
  {Qi}}\ and\ \bibinfo {author} {\bibfnamefont {S.-C.}\ \bibnamefont {Zhang}},\
  }\Doi {10.1103/RevModPhys.83.1057} {\bibfield  {journal} {\bibinfo  {journal}
  {Rev. Mod. Phys.},\ }\textbf {\bibinfo {volume} {83}},\ \bibinfo {pages}
  {1057} (\bibinfo {year} {2011})}\BibitemShut {NoStop}%
\bibitem [{\citenamefont {Ando}(2013)}]{Ando_JPSJ_2013}%
  \BibitemOpen
  \bibfield  {author} {\bibinfo {author} {\bibfnamefont {Y.}~\bibnamefont
  {Ando}},\ }\href@noop {} {\bibfield  {journal} {\bibinfo  {journal} {Journal
  of the Physical Society of Japan},\ }\textbf {\bibinfo {volume} {82}},\
  \bibinfo {pages} {102001} (\bibinfo {year} {2013})}\BibitemShut {NoStop}%
\bibitem [{\citenamefont {Klitzing}\ \emph {et~al.}(1980)\citenamefont
  {Klitzing}, \citenamefont {Dorda},\ and\ \citenamefont
  {Pepper}}]{Klitzing_PRL_1980}%
  \BibitemOpen
  \bibfield  {author} {\bibinfo {author} {\bibfnamefont {K.~v.}\ \bibnamefont
  {Klitzing}}, \bibinfo {author} {\bibfnamefont {G.}~\bibnamefont {Dorda}}, \
  and\ \bibinfo {author} {\bibfnamefont {M.}~\bibnamefont {Pepper}},\ }\Doi
  {10.1103/PhysRevLett.45.494} {\bibfield  {journal} {\bibinfo  {journal}
  {Phys. Rev. Lett.},\ }\textbf {\bibinfo {volume} {45}},\ \bibinfo {pages}
  {494} (\bibinfo {year} {1980})}\BibitemShut {NoStop}%
\bibitem [{\citenamefont {Laughlin}(1981)}]{Laughlin_PRB_1981}%
  \BibitemOpen
  \bibfield  {author} {\bibinfo {author} {\bibfnamefont {R.~B.}\ \bibnamefont
  {Laughlin}},\ }\Doi {10.1103/PhysRevB.23.5632} {\bibfield  {journal}
  {\bibinfo  {journal} {Phys. Rev. B},\ }\textbf {\bibinfo {volume} {23}},\
  \bibinfo {pages} {5632} (\bibinfo {year} {1981})}\BibitemShut {NoStop}%
\bibitem [{\citenamefont {Thouless}\ \emph {et~al.}(1982)\citenamefont
  {Thouless}, \citenamefont {Kohmoto}, \citenamefont {Nightingale},\ and\
  \citenamefont {den Nijs}}]{Thouless_PRL_1982}%
  \BibitemOpen
  \bibfield  {author} {\bibinfo {author} {\bibfnamefont {D.~J.}\ \bibnamefont
  {Thouless}}, \bibinfo {author} {\bibfnamefont {M.}~\bibnamefont {Kohmoto}},
  \bibinfo {author} {\bibfnamefont {M.~P.}\ \bibnamefont {Nightingale}}, \ and\
  \bibinfo {author} {\bibfnamefont {M.}~\bibnamefont {den Nijs}},\ }\Doi
  {10.1103/PhysRevLett.49.405} {\bibfield  {journal} {\bibinfo  {journal}
  {Phys. Rev. Lett.},\ }\textbf {\bibinfo {volume} {49}},\ \bibinfo {pages}
  {405} (\bibinfo {year} {1982})}\BibitemShut {NoStop}%
\bibitem [{\citenamefont {Haldane}(1988)}]{Haldane_PRL_1988}%
  \BibitemOpen
  \bibfield  {author} {\bibinfo {author} {\bibfnamefont {F.~D.~M.}\
  \bibnamefont {Haldane}},\ }\Doi {10.1103/PhysRevLett.61.2015} {\bibfield
  {journal} {\bibinfo  {journal} {Phys. Rev. Lett.},\ }\textbf {\bibinfo
  {volume} {61}},\ \bibinfo {pages} {2015} (\bibinfo {year}
  {1988})}\BibitemShut {NoStop}%
\bibitem [{\citenamefont {Murakami}\ \emph {et~al.}(2004)\citenamefont
  {Murakami}, \citenamefont {Nagaosa},\ and\ \citenamefont
  {Zhang}}]{Murakami_PRL_2004}%
  \BibitemOpen
  \bibfield  {author} {\bibinfo {author} {\bibfnamefont {S.}~\bibnamefont
  {Murakami}}, \bibinfo {author} {\bibfnamefont {N.}~\bibnamefont {Nagaosa}}, \
  and\ \bibinfo {author} {\bibfnamefont {S.-C.}\ \bibnamefont {Zhang}},\ }\Doi
  {10.1103/PhysRevLett.93.156804} {\bibfield  {journal} {\bibinfo  {journal}
  {Phys. Rev. Lett.},\ }\textbf {\bibinfo {volume} {93}},\ \bibinfo {pages}
  {156804} (\bibinfo {year} {2004})}\BibitemShut {NoStop}%
\bibitem [{\citenamefont {Kane}\ and\ \citenamefont
  {Mele}(2005){\natexlab{a}}}]{Kane_PRL_2005a}%
  \BibitemOpen
  \bibfield  {author} {\bibinfo {author} {\bibfnamefont {C.~L.}\ \bibnamefont
  {Kane}}\ and\ \bibinfo {author} {\bibfnamefont {E.~J.}\ \bibnamefont
  {Mele}},\ }\Doi {10.1103/PhysRevLett.95.146802} {\bibfield  {journal}
  {\bibinfo  {journal} {Phys. Rev. Lett.},\ }\textbf {\bibinfo {volume} {95}},\
  \bibinfo {pages} {146802} (\bibinfo {year} {2005}{\natexlab{a}})}\BibitemShut
  {NoStop}%
\bibitem [{\citenamefont {Kane}\ and\ \citenamefont
  {Mele}(2005){\natexlab{b}}}]{Kane_PRL_2005b}%
  \BibitemOpen
  \bibfield  {author} {\bibinfo {author} {\bibfnamefont {C.~L.}\ \bibnamefont
  {Kane}}\ and\ \bibinfo {author} {\bibfnamefont {E.~J.}\ \bibnamefont
  {Mele}},\ }\Doi {10.1103/PhysRevLett.95.226801} {\bibfield  {journal}
  {\bibinfo  {journal} {Phys. Rev. Lett.},\ }\textbf {\bibinfo {volume} {95}},\
  \bibinfo {pages} {226801} (\bibinfo {year} {2005}{\natexlab{b}})}\BibitemShut
  {NoStop}%
\bibitem [{\citenamefont {Bernevig}\ and\ \citenamefont
  {Zhang}(2006)}]{Bernevig_PRL_2006}%
  \BibitemOpen
  \bibfield  {author} {\bibinfo {author} {\bibfnamefont {B.~A.}\ \bibnamefont
  {Bernevig}}\ and\ \bibinfo {author} {\bibfnamefont {S.-C.}\ \bibnamefont
  {Zhang}},\ }\Doi {10.1103/PhysRevLett.96.106802} {\bibfield  {journal}
  {\bibinfo  {journal} {Phys. Rev. Lett.},\ }\textbf {\bibinfo {volume} {96}},\
  \bibinfo {pages} {106802} (\bibinfo {year} {2006})}\BibitemShut {NoStop}%
\bibitem [{\citenamefont {Bernevig}\ \emph {et~al.}(2006)\citenamefont
  {Bernevig}, \citenamefont {Hughes},\ and\ \citenamefont
  {Zhang}}]{Bernevig_Science_2006}%
  \BibitemOpen
  \bibfield  {author} {\bibinfo {author} {\bibfnamefont {B.~A.}\ \bibnamefont
  {Bernevig}}, \bibinfo {author} {\bibfnamefont {T.~L.}\ \bibnamefont
  {Hughes}}, \ and\ \bibinfo {author} {\bibfnamefont {S.-C.}\ \bibnamefont
  {Zhang}},\ }\Doi {10.1126/science.1133734} {\bibfield  {journal} {\bibinfo
  {journal} {Science},\ }\textbf {\bibinfo {volume} {314}},\ \bibinfo {pages}
  {1757} (\bibinfo {year} {2006})},\ ISSN \bibinfo {issn} {0036-8075},\ \Eprint
  {http://arxiv.org/abs/http://science.sciencemag.org/content/314/5806/1757.full.pdf}
  {http://science.sciencemag.org/content/314/5806/1757.full.pdf} \BibitemShut
  {NoStop}%
\bibitem [{\citenamefont {K\"{o}nig}\ \emph {et~al.}(2007)\citenamefont
  {K\"{o}nig}, \citenamefont {Wiedmann}, \citenamefont {BrÃ¼ne},
  \citenamefont {Roth}, \citenamefont {Buhmann}, \citenamefont {Molenkamp},
  \citenamefont {Qi},\ and\ \citenamefont {Zhang}}]{Konig_Science_2007a}%
  \BibitemOpen
  \bibfield  {author} {\bibinfo {author} {\bibfnamefont {M.}~\bibnamefont
  {K\"{o}nig}}, \bibinfo {author} {\bibfnamefont {S.}~\bibnamefont {Wiedmann}},
  \bibinfo {author} {\bibfnamefont {C.}~\bibnamefont {BrÃ¼ne}}, \bibinfo
  {author} {\bibfnamefont {A.}~\bibnamefont {Roth}}, \bibinfo {author}
  {\bibfnamefont {H.}~\bibnamefont {Buhmann}}, \bibinfo {author} {\bibfnamefont
  {L.~W.}\ \bibnamefont {Molenkamp}}, \bibinfo {author} {\bibfnamefont {X.-L.}\
  \bibnamefont {Qi}}, \ and\ \bibinfo {author} {\bibfnamefont {S.-C.}\
  \bibnamefont {Zhang}},\ }\Doi {10.1126/science.1148047} {\bibfield  {journal}
  {\bibinfo  {journal} {Science},\ }\textbf {\bibinfo {volume} {318}},\
  \bibinfo {pages} {766} (\bibinfo {year} {2007})}\BibitemShut {NoStop}%
\bibitem [{\citenamefont {Fu}\ \emph {et~al.}(2007)\citenamefont {Fu},
  \citenamefont {Kane},\ and\ \citenamefont {Mele}}]{Fu_PRL_2007}%
  \BibitemOpen
  \bibfield  {author} {\bibinfo {author} {\bibfnamefont {L.}~\bibnamefont
  {Fu}}, \bibinfo {author} {\bibfnamefont {C.~L.}\ \bibnamefont {Kane}}, \ and\
  \bibinfo {author} {\bibfnamefont {E.~J.}\ \bibnamefont {Mele}},\ }\Doi
  {10.1103/PhysRevLett.98.106803} {\bibfield  {journal} {\bibinfo  {journal}
  {Phys. Rev. Lett.},\ }\textbf {\bibinfo {volume} {98}},\ \bibinfo {pages}
  {106803} (\bibinfo {year} {2007})}\BibitemShut {NoStop}%
\bibitem [{\citenamefont {Moore}\ and\ \citenamefont
  {Balents}(2007)}]{Moore_PRB_2007}%
  \BibitemOpen
  \bibfield  {author} {\bibinfo {author} {\bibfnamefont {J.~E.}\ \bibnamefont
  {Moore}}\ and\ \bibinfo {author} {\bibfnamefont {L.}~\bibnamefont
  {Balents}},\ }\Doi {10.1103/PhysRevB.75.121306} {\bibfield  {journal}
  {\bibinfo  {journal} {Phys. Rev. B},\ }\textbf {\bibinfo {volume} {75}},\
  \bibinfo {pages} {121306} (\bibinfo {year} {2007})}\BibitemShut {NoStop}%
\bibitem [{\citenamefont {Roy}(2009)}]{Roy_PRB_2009b}%
  \BibitemOpen
  \bibfield  {author} {\bibinfo {author} {\bibfnamefont {R.}~\bibnamefont
  {Roy}},\ }\Doi {10.1103/PhysRevB.79.195322} {\bibfield  {journal} {\bibinfo
  {journal} {Phys. Rev. B},\ }\textbf {\bibinfo {volume} {79}},\ \bibinfo
  {pages} {195322} (\bibinfo {year} {2009})}\BibitemShut {NoStop}%
\bibitem [{\citenamefont {Hsieh}\ \emph {et~al.}(2008)\citenamefont {Hsieh},
  \citenamefont {Qian}, \citenamefont {Wray}, \citenamefont {Xia},
  \citenamefont {Hor}, \citenamefont {Cava},\ and\ \citenamefont
  {Hasan}}]{Hsieh_Nature_2008}%
  \BibitemOpen
  \bibfield  {author} {\bibinfo {author} {\bibfnamefont {D.}~\bibnamefont
  {Hsieh}}, \bibinfo {author} {\bibfnamefont {D.}~\bibnamefont {Qian}},
  \bibinfo {author} {\bibfnamefont {L.}~\bibnamefont {Wray}}, \bibinfo {author}
  {\bibfnamefont {Y.}~\bibnamefont {Xia}}, \bibinfo {author} {\bibfnamefont
  {Y.~S.}\ \bibnamefont {Hor}}, \bibinfo {author} {\bibfnamefont {R.~J.}\
  \bibnamefont {Cava}}, \ and\ \bibinfo {author} {\bibfnamefont {M.~Z.}\
  \bibnamefont {Hasan}},\ }\Doi {10.1038/nature06843} {\bibfield  {journal}
  {\bibinfo  {journal} {Nature},\ }\textbf {\bibinfo {volume} {452}},\ \bibinfo
  {pages} {970} (\bibinfo {year} {2008})},\ ISSN \bibinfo {issn}
  {0028-0836}\BibitemShut {NoStop}%
\bibitem [{\citenamefont {Qi}\ \emph {et~al.}(2008)\citenamefont {Qi},
  \citenamefont {Hughes},\ and\ \citenamefont {Zhang}}]{Qi_PRB_2008}%
  \BibitemOpen
  \bibfield  {author} {\bibinfo {author} {\bibfnamefont {X.-L.}\ \bibnamefont
  {Qi}}, \bibinfo {author} {\bibfnamefont {T.~L.}\ \bibnamefont {Hughes}}, \
  and\ \bibinfo {author} {\bibfnamefont {S.-C.}\ \bibnamefont {Zhang}},\ }\Doi
  {10.1103/PhysRevB.78.195424} {\bibfield  {journal} {\bibinfo  {journal}
  {Phys. Rev. B},\ }\textbf {\bibinfo {volume} {78}},\ \bibinfo {pages}
  {195424} (\bibinfo {year} {2008})}\BibitemShut {NoStop}%
\bibitem [{\citenamefont {Schnyder}\ \emph {et~al.}(2008)\citenamefont
  {Schnyder}, \citenamefont {Ryu}, \citenamefont {Furusaki},\ and\
  \citenamefont {Ludwig}}]{Schnyder_PRB_2008}%
  \BibitemOpen
  \bibfield  {author} {\bibinfo {author} {\bibfnamefont {A.~P.}\ \bibnamefont
  {Schnyder}}, \bibinfo {author} {\bibfnamefont {S.}~\bibnamefont {Ryu}},
  \bibinfo {author} {\bibfnamefont {A.}~\bibnamefont {Furusaki}}, \ and\
  \bibinfo {author} {\bibfnamefont {A.~W.~W.}\ \bibnamefont {Ludwig}},\ }\Doi
  {10.1103/PhysRevB.78.195125} {\bibfield  {journal} {\bibinfo  {journal}
  {Phys. Rev. B},\ }\textbf {\bibinfo {volume} {78}},\ \bibinfo {pages}
  {195125} (\bibinfo {year} {2008})}\BibitemShut {NoStop}%
\bibitem [{\citenamefont {Ryu}\ \emph {et~al.}(2010)\citenamefont {Ryu},
  \citenamefont {Schnyder}, \citenamefont {Furusaki},\ and\ \citenamefont
  {Ludwig}}]{Ryu_NJP_2010}%
  \BibitemOpen
  \bibfield  {author} {\bibinfo {author} {\bibfnamefont {S.}~\bibnamefont
  {Ryu}}, \bibinfo {author} {\bibfnamefont {A.~P.}\ \bibnamefont {Schnyder}},
  \bibinfo {author} {\bibfnamefont {A.}~\bibnamefont {Furusaki}}, \ and\
  \bibinfo {author} {\bibfnamefont {A.~W.~W.}\ \bibnamefont {Ludwig}},\ }\href
  {http://stacks.iop.org/1367-2630/12/i=6/a=065010} {\bibfield  {journal}
  {\bibinfo  {journal} {New Journal of Physics},\ }\textbf {\bibinfo {volume}
  {12}},\ \bibinfo {pages} {065010} (\bibinfo {year} {2010})}\BibitemShut
  {NoStop}%
\bibitem [{\citenamefont {Kitaev}(2009)}]{Kitaev_AIP_2009}%
  \BibitemOpen
  \bibfield  {author} {\bibinfo {author} {\bibfnamefont {A.}~\bibnamefont
  {Kitaev}},\ }\Doi {10.1063/1.3149495} {\bibfield  {journal} {\bibinfo
  {journal} {AIP Conference Proceedings},\ }\textbf {\bibinfo {volume}
  {1134}},\ \bibinfo {pages} {22} (\bibinfo {year} {2009})},\ \Eprint
  {http://arxiv.org/abs/http://aip.scitation.org/doi/pdf/10.1063/1.3149495}
  {http://aip.scitation.org/doi/pdf/10.1063/1.3149495} \BibitemShut {NoStop}%
\bibitem [{\citenamefont {Altland}\ and\ \citenamefont
  {Zirnbauer}(1997)}]{Altland_PRB_1997}%
  \BibitemOpen
  \bibfield  {author} {\bibinfo {author} {\bibfnamefont {A.}~\bibnamefont
  {Altland}}\ and\ \bibinfo {author} {\bibfnamefont {M.~R.}\ \bibnamefont
  {Zirnbauer}},\ }\Doi {10.1103/PhysRevB.55.1142} {\bibfield  {journal}
  {\bibinfo  {journal} {Phys. Rev. B},\ }\textbf {\bibinfo {volume} {55}},\
  \bibinfo {pages} {1142} (\bibinfo {year} {1997})}\BibitemShut {NoStop}%
\bibitem [{\citenamefont {Kitaev}(2001)}]{Kitaev_PU_2001}%
  \BibitemOpen
  \bibfield  {author} {\bibinfo {author} {\bibfnamefont {A.~Y.}\ \bibnamefont
  {Kitaev}},\ }\href {http://stacks.iop.org/1063-7869/44/i=10S/a=S29}
  {\bibfield  {journal} {\bibinfo  {journal} {Physics-Uspekhi},\ }\textbf
  {\bibinfo {volume} {44}},\ \bibinfo {pages} {131} (\bibinfo {year}
  {2001})}\BibitemShut {NoStop}%
\bibitem [{\citenamefont {Das}\ \emph {et~al.}(2012)\citenamefont {Das},
  \citenamefont {Ronen}, \citenamefont {Most}, \citenamefont {Oreg},
  \citenamefont {Heiblum},\ and\ \citenamefont {Shtrikman}}]{Das_NatPhys_2012}%
  \BibitemOpen
  \bibfield  {author} {\bibinfo {author} {\bibfnamefont {A.}~\bibnamefont
  {Das}}, \bibinfo {author} {\bibfnamefont {Y.}~\bibnamefont {Ronen}}, \bibinfo
  {author} {\bibfnamefont {Y.}~\bibnamefont {Most}}, \bibinfo {author}
  {\bibfnamefont {Y.}~\bibnamefont {Oreg}}, \bibinfo {author} {\bibfnamefont
  {M.}~\bibnamefont {Heiblum}}, \ and\ \bibinfo {author} {\bibfnamefont
  {H.}~\bibnamefont {Shtrikman}},\ }\href@noop {} {\bibfield  {journal}
  {\bibinfo  {journal} {Nature Physics},\ }\textbf {\bibinfo {volume} {8}},\
  \bibinfo {pages} {887} (\bibinfo {year} {2012})}\BibitemShut {NoStop}%
\bibitem [{\citenamefont {Chang}\ \emph {et~al.}(2013)\citenamefont {Chang},
  \citenamefont {Zhang}, \citenamefont {Feng}, \citenamefont {Shen},
  \citenamefont {Zhang}, \citenamefont {Guo}, \citenamefont {Li}, \citenamefont
  {Ou}, \citenamefont {Wei}, \citenamefont {Wang} \emph
  {et~al.}}]{Chang_Science_2013}%
  \BibitemOpen
  \bibfield  {author} {\bibinfo {author} {\bibfnamefont {C.-Z.}\ \bibnamefont
  {Chang}}, \bibinfo {author} {\bibfnamefont {J.}~\bibnamefont {Zhang}},
  \bibinfo {author} {\bibfnamefont {X.}~\bibnamefont {Feng}}, \bibinfo {author}
  {\bibfnamefont {J.}~\bibnamefont {Shen}}, \bibinfo {author} {\bibfnamefont
  {Z.}~\bibnamefont {Zhang}}, \bibinfo {author} {\bibfnamefont
  {M.}~\bibnamefont {Guo}}, \bibinfo {author} {\bibfnamefont {K.}~\bibnamefont
  {Li}}, \bibinfo {author} {\bibfnamefont {Y.}~\bibnamefont {Ou}}, \bibinfo
  {author} {\bibfnamefont {P.}~\bibnamefont {Wei}}, \bibinfo {author}
  {\bibfnamefont {L.-L.}\ \bibnamefont {Wang}},  \emph {et~al.},\ }\href@noop
  {} {\bibfield  {journal} {\bibinfo  {journal} {Science},\ }\textbf {\bibinfo
  {volume} {340}},\ \bibinfo {pages} {167} (\bibinfo {year}
  {2013})}\BibitemShut {NoStop}%
\bibitem [{\citenamefont {Nadj-Perge}\ \emph {et~al.}(2014)\citenamefont
  {Nadj-Perge}, \citenamefont {Drozdov}, \citenamefont {Li}, \citenamefont
  {Chen}, \citenamefont {Jeon}, \citenamefont {Seo}, \citenamefont {MacDonald},
  \citenamefont {Bernevig},\ and\ \citenamefont {Yazdani}}]{Nadj_Science_2014}%
  \BibitemOpen
  \bibfield  {author} {\bibinfo {author} {\bibfnamefont {S.}~\bibnamefont
  {Nadj-Perge}}, \bibinfo {author} {\bibfnamefont {I.~K.}\ \bibnamefont
  {Drozdov}}, \bibinfo {author} {\bibfnamefont {J.}~\bibnamefont {Li}},
  \bibinfo {author} {\bibfnamefont {H.}~\bibnamefont {Chen}}, \bibinfo {author}
  {\bibfnamefont {S.}~\bibnamefont {Jeon}}, \bibinfo {author} {\bibfnamefont
  {J.}~\bibnamefont {Seo}}, \bibinfo {author} {\bibfnamefont {A.~H.}\
  \bibnamefont {MacDonald}}, \bibinfo {author} {\bibfnamefont {B.~A.}\
  \bibnamefont {Bernevig}}, \ and\ \bibinfo {author} {\bibfnamefont
  {A.}~\bibnamefont {Yazdani}},\ }\href@noop {} {\bibfield  {journal} {\bibinfo
   {journal} {Science},\ }\textbf {\bibinfo {volume} {346}},\ \bibinfo {pages}
  {602} (\bibinfo {year} {2014})}\BibitemShut {NoStop}%
\bibitem [{\citenamefont {Jotzu}\ \emph {et~al.}(2014)\citenamefont {Jotzu},
  \citenamefont {Messer}, \citenamefont {Desbuquois}, \citenamefont {Lebrat},
  \citenamefont {Uehlinger}, \citenamefont {Greif},\ and\ \citenamefont
  {Esslinger}}]{Jotzu_Nature_2014}%
  \BibitemOpen
  \bibfield  {author} {\bibinfo {author} {\bibfnamefont {G.}~\bibnamefont
  {Jotzu}}, \bibinfo {author} {\bibfnamefont {M.}~\bibnamefont {Messer}},
  \bibinfo {author} {\bibfnamefont {R.}~\bibnamefont {Desbuquois}}, \bibinfo
  {author} {\bibfnamefont {M.}~\bibnamefont {Lebrat}}, \bibinfo {author}
  {\bibfnamefont {T.}~\bibnamefont {Uehlinger}}, \bibinfo {author}
  {\bibfnamefont {D.}~\bibnamefont {Greif}}, \ and\ \bibinfo {author}
  {\bibfnamefont {T.}~\bibnamefont {Esslinger}},\ }\href@noop {} {\bibfield
  {journal} {\bibinfo  {journal} {Nature},\ }\textbf {\bibinfo {volume}
  {515}},\ \bibinfo {pages} {237} (\bibinfo {year} {2014})}\BibitemShut
  {NoStop}%
\bibitem [{\citenamefont {Kobayashi}\ \emph {et~al.}(2013)\citenamefont
  {Kobayashi}, \citenamefont {Ohtsuki},\ and\ \citenamefont
  {Imura}}]{Kobayashi_PRL_2013}%
  \BibitemOpen
  \bibfield  {author} {\bibinfo {author} {\bibfnamefont {K.}~\bibnamefont
  {Kobayashi}}, \bibinfo {author} {\bibfnamefont {T.}~\bibnamefont {Ohtsuki}},
  \ and\ \bibinfo {author} {\bibfnamefont {K.-I.}\ \bibnamefont {Imura}},\
  }\Doi {10.1103/PhysRevLett.110.236803} {\bibfield  {journal} {\bibinfo
  {journal} {Phys. Rev. Lett.},\ }\textbf {\bibinfo {volume} {110}},\ \bibinfo
  {pages} {236803} (\bibinfo {year} {2013})}\BibitemShut {NoStop}%
\bibitem [{\citenamefont {Diez}\ \emph {et~al.}(2014)\citenamefont {Diez},
  \citenamefont {Fulga}, \citenamefont {Pikulin}, \citenamefont {Tworzydło},\
  and\ \citenamefont {Beenakker}}]{Diez_NJP_2014}%
  \BibitemOpen
  \bibfield  {author} {\bibinfo {author} {\bibfnamefont {M.}~\bibnamefont
  {Diez}}, \bibinfo {author} {\bibfnamefont {I.~C.}\ \bibnamefont {Fulga}},
  \bibinfo {author} {\bibfnamefont {D.~I.}\ \bibnamefont {Pikulin}}, \bibinfo
  {author} {\bibfnamefont {J.}~\bibnamefont {Tworzydło}}, \ and\ \bibinfo
  {author} {\bibfnamefont {C.~W.~J.}\ \bibnamefont {Beenakker}},\ }\href
  {http://stacks.iop.org/1367-2630/16/i=6/a=063049} {\bibfield  {journal}
  {\bibinfo  {journal} {New Journal of Physics},\ }\textbf {\bibinfo {volume}
  {16}},\ \bibinfo {pages} {063049} (\bibinfo {year} {2014})}\BibitemShut
  {NoStop}%
\bibitem [{\citenamefont {Li}\ \emph {et~al.}(2009)\citenamefont {Li},
  \citenamefont {Chu}, \citenamefont {Jain},\ and\ \citenamefont
  {Shen}}]{Li_PRL_2009}%
  \BibitemOpen
  \bibfield  {author} {\bibinfo {author} {\bibfnamefont {J.}~\bibnamefont
  {Li}}, \bibinfo {author} {\bibfnamefont {R.-L.}\ \bibnamefont {Chu}},
  \bibinfo {author} {\bibfnamefont {J.~K.}\ \bibnamefont {Jain}}, \ and\
  \bibinfo {author} {\bibfnamefont {S.-Q.}\ \bibnamefont {Shen}},\ }\Doi
  {10.1103/PhysRevLett.102.136806} {\bibfield  {journal} {\bibinfo  {journal}
  {Phys. Rev. Lett.},\ }\textbf {\bibinfo {volume} {102}},\ \bibinfo {pages}
  {136806} (\bibinfo {year} {2009})}\BibitemShut {NoStop}%
\bibitem [{\citenamefont {Fulga}\ \emph {et~al.}(2014)\citenamefont {Fulga},
  \citenamefont {van Heck}, \citenamefont {Edge},\ and\ \citenamefont
  {Akhmerov}}]{Fulga_PRB_2014}%
  \BibitemOpen
  \bibfield  {author} {\bibinfo {author} {\bibfnamefont {I.~C.}\ \bibnamefont
  {Fulga}}, \bibinfo {author} {\bibfnamefont {B.}~\bibnamefont {van Heck}},
  \bibinfo {author} {\bibfnamefont {J.~M.}\ \bibnamefont {Edge}}, \ and\
  \bibinfo {author} {\bibfnamefont {A.~R.}\ \bibnamefont {Akhmerov}},\ }\Doi
  {10.1103/PhysRevB.89.155424} {\bibfield  {journal} {\bibinfo  {journal}
  {Phys. Rev. B},\ }\textbf {\bibinfo {volume} {89}},\ \bibinfo {pages}
  {155424} (\bibinfo {year} {2014})}\BibitemShut {NoStop}%
\bibitem [{\citenamefont {Ringel}\ \emph {et~al.}(2012)\citenamefont {Ringel},
  \citenamefont {Kraus},\ and\ \citenamefont {Stern}}]{Ringel_PRB_2012}%
  \BibitemOpen
  \bibfield  {author} {\bibinfo {author} {\bibfnamefont {Z.}~\bibnamefont
  {Ringel}}, \bibinfo {author} {\bibfnamefont {Y.~E.}\ \bibnamefont {Kraus}}, \
  and\ \bibinfo {author} {\bibfnamefont {A.}~\bibnamefont {Stern}},\ }\Doi
  {10.1103/PhysRevB.86.045102} {\bibfield  {journal} {\bibinfo  {journal}
  {Phys. Rev. B},\ }\textbf {\bibinfo {volume} {86}},\ \bibinfo {pages}
  {045102} (\bibinfo {year} {2012})}\BibitemShut {NoStop}%
\bibitem [{\citenamefont {Kraus}\ \emph {et~al.}(2012)\citenamefont {Kraus},
  \citenamefont {Lahini}, \citenamefont {Ringel}, \citenamefont {Verbin},\ and\
  \citenamefont {Zilberberg}}]{Kraus_PRL_2012}%
  \BibitemOpen
  \bibfield  {author} {\bibinfo {author} {\bibfnamefont {Y.~E.}\ \bibnamefont
  {Kraus}}, \bibinfo {author} {\bibfnamefont {Y.}~\bibnamefont {Lahini}},
  \bibinfo {author} {\bibfnamefont {Z.}~\bibnamefont {Ringel}}, \bibinfo
  {author} {\bibfnamefont {M.}~\bibnamefont {Verbin}}, \ and\ \bibinfo {author}
  {\bibfnamefont {O.}~\bibnamefont {Zilberberg}},\ }\Doi
  {10.1103/PhysRevLett.109.106402} {\bibfield  {journal} {\bibinfo  {journal}
  {Phys. Rev. Lett.},\ }\textbf {\bibinfo {volume} {109}},\ \bibinfo {pages}
  {106402} (\bibinfo {year} {2012})}\BibitemShut {NoStop}%
\bibitem [{\citenamefont {Fulga}\ \emph {et~al.}(2016)\citenamefont {Fulga},
  \citenamefont {Pikulin},\ and\ \citenamefont {Loring}}]{Fulga_PRL_2016}%
  \BibitemOpen
  \bibfield  {author} {\bibinfo {author} {\bibfnamefont {I.~C.}\ \bibnamefont
  {Fulga}}, \bibinfo {author} {\bibfnamefont {D.~I.}\ \bibnamefont {Pikulin}},
  \ and\ \bibinfo {author} {\bibfnamefont {T.~A.}\ \bibnamefont {Loring}},\
  }\Doi {10.1103/PhysRevLett.116.257002} {\bibfield  {journal} {\bibinfo
  {journal} {Phys. Rev. Lett.},\ }\textbf {\bibinfo {volume} {116}},\ \bibinfo
  {pages} {257002} (\bibinfo {year} {2016})}\BibitemShut {NoStop}%
\bibitem [{\citenamefont {Bandres}\ \emph {et~al.}(2016)\citenamefont
  {Bandres}, \citenamefont {Rechtsman},\ and\ \citenamefont
  {Segev}}]{Bandres_PRX_2016}%
  \BibitemOpen
  \bibfield  {author} {\bibinfo {author} {\bibfnamefont {M.~A.}\ \bibnamefont
  {Bandres}}, \bibinfo {author} {\bibfnamefont {M.~C.}\ \bibnamefont
  {Rechtsman}}, \ and\ \bibinfo {author} {\bibfnamefont {M.}~\bibnamefont
  {Segev}},\ }\Doi {10.1103/PhysRevX.6.011016} {\bibfield  {journal} {\bibinfo
  {journal} {Phys. Rev. X},\ }\textbf {\bibinfo {volume} {6}},\ \bibinfo
  {pages} {011016} (\bibinfo {year} {2016})}\BibitemShut {NoStop}%
\bibitem [{\citenamefont {Christ}\ \emph {et~al.}(1982)\citenamefont {Christ},
  \citenamefont {Friedberg},\ and\ \citenamefont {Lee}}]{Christ_NPB_1982}%
  \BibitemOpen
  \bibfield  {author} {\bibinfo {author} {\bibfnamefont {N.}~\bibnamefont
  {Christ}}, \bibinfo {author} {\bibfnamefont {R.}~\bibnamefont {Friedberg}}, \
  and\ \bibinfo {author} {\bibfnamefont {T.}~\bibnamefont {Lee}},\ }\Doi
  {10.1016/0550-3213(82)90222-X} {\bibfield  {journal} {\bibinfo  {journal}
  {Nuclear Physics B},\ }\textbf {\bibinfo {volume} {202}},\ \bibinfo {pages}
  {89 } (\bibinfo {year} {1982})},\ ISSN \bibinfo {issn}
  {0550-3213}\BibitemShut {NoStop}%
\bibitem [{\citenamefont {{Loring, T. A.}}\ and\ \citenamefont {{Hastings, M.
  B.}}(2010)}]{Loring_EPL_2010}%
  \BibitemOpen
  \bibfield  {author} {\bibinfo {author} {\bibnamefont {{Loring, T. A.}}}\ and\
  \bibinfo {author} {\bibnamefont {{Hastings, M. B.}}},\ }\Doi
  {10.1209/0295-5075/92/67004} {\bibfield  {journal} {\bibinfo  {journal}
  {EPL},\ }\textbf {\bibinfo {volume} {92}},\ \bibinfo {pages} {67004}
  (\bibinfo {year} {2010})}\BibitemShut {NoStop}%
\bibitem [{\citenamefont {Bernevig}\ and\ \citenamefont
  {Hughes}(2013)}]{Bernevig_Book_2013}%
  \BibitemOpen
  \bibfield  {author} {\bibinfo {author} {\bibfnamefont {B.~A.}\ \bibnamefont
  {Bernevig}}\ and\ \bibinfo {author} {\bibfnamefont {T.~L.}\ \bibnamefont
  {Hughes}},\ }\href@noop {} {\emph {\bibinfo {title} {Topological insulators
  and topological superconductors}}}\ (\bibinfo  {publisher} {Princeton
  University Press},\ \bibinfo {year} {2013})\BibitemShut {NoStop}%
\bibitem [{\citenamefont {Roy}(2006)}]{Roy_arXiv_2006}%
  \BibitemOpen
  \bibfield  {author} {\bibinfo {author} {\bibfnamefont {R.}~\bibnamefont
  {Roy}},\ }\href@noop {} {\bibfield  {journal} {\bibinfo  {journal} {arXiv
  preprint cond-mat/0608064}} (\bibinfo {year} {2006})}\BibitemShut {NoStop}%
\bibitem [{\citenamefont {Qi}\ \emph {et~al.}(2009)\citenamefont {Qi},
  \citenamefont {Hughes}, \citenamefont {Raghu},\ and\ \citenamefont
  {Zhang}}]{Qi_PRL_2009}%
  \BibitemOpen
  \bibfield  {author} {\bibinfo {author} {\bibfnamefont {X.-L.}\ \bibnamefont
  {Qi}}, \bibinfo {author} {\bibfnamefont {T.~L.}\ \bibnamefont {Hughes}},
  \bibinfo {author} {\bibfnamefont {S.}~\bibnamefont {Raghu}}, \ and\ \bibinfo
  {author} {\bibfnamefont {S.-C.}\ \bibnamefont {Zhang}},\ }\Doi
  {10.1103/PhysRevLett.102.187001} {\bibfield  {journal} {\bibinfo  {journal}
  {Phys. Rev. Lett.},\ }\textbf {\bibinfo {volume} {102}},\ \bibinfo {pages}
  {187001} (\bibinfo {year} {2009})}\BibitemShut {NoStop}%
\bibitem [{\citenamefont {Senthil}\ \emph {et~al.}(1999)\citenamefont
  {Senthil}, \citenamefont {Marston},\ and\ \citenamefont
  {Fisher}}]{Senthil_PRB_1999}%
  \BibitemOpen
  \bibfield  {author} {\bibinfo {author} {\bibfnamefont {T.}~\bibnamefont
  {Senthil}}, \bibinfo {author} {\bibfnamefont {J.~B.}\ \bibnamefont
  {Marston}}, \ and\ \bibinfo {author} {\bibfnamefont {M.~P.~A.}\ \bibnamefont
  {Fisher}},\ }\Doi {10.1103/PhysRevB.60.4245} {\bibfield  {journal} {\bibinfo
  {journal} {Phys. Rev. B},\ }\textbf {\bibinfo {volume} {60}},\ \bibinfo
  {pages} {4245} (\bibinfo {year} {1999})}\BibitemShut {NoStop}%
\bibitem [{\citenamefont {Chern}(2016)}]{Chern_AIP_2016}%
  \BibitemOpen
  \bibfield  {author} {\bibinfo {author} {\bibfnamefont {T.}~\bibnamefont
  {Chern}},\ }\Doi {10.1063/1.4961462} {\bibfield  {journal} {\bibinfo
  {journal} {AIP Advances},\ }\textbf {\bibinfo {volume} {6}} (\bibinfo {year}
  {2016})},\ \doi {10.1063/1.4961462}\BibitemShut {NoStop}%
\bibitem [{\citenamefont {Datta}(1997)}]{Datta_Book_1997}%
  \BibitemOpen
  \bibfield  {author} {\bibinfo {author} {\bibfnamefont {S.}~\bibnamefont
  {Datta}},\ }\href@noop {} {\emph {\bibinfo {title} {Electronic transport in
  mesoscopic systems}}}\ (\bibinfo  {publisher} {Cambridge university press},\
  \bibinfo {year} {1997})\BibitemShut {NoStop}%
\bibitem [{\citenamefont {Scappucci}\ \emph {et~al.}(2009)\citenamefont
  {Scappucci}, \citenamefont {Capellini}, \citenamefont {Lee},\ and\
  \citenamefont {Simmons}}]{Scappucci_APL_2009}%
  \BibitemOpen
  \bibfield  {author} {\bibinfo {author} {\bibfnamefont {G.}~\bibnamefont
  {Scappucci}}, \bibinfo {author} {\bibfnamefont {G.}~\bibnamefont
  {Capellini}}, \bibinfo {author} {\bibfnamefont {W.~C.~T.}\ \bibnamefont
  {Lee}}, \ and\ \bibinfo {author} {\bibfnamefont {M.~Y.}\ \bibnamefont
  {Simmons}},\ }\Doi {10.1063/1.3123391} {\bibfield  {journal} {\bibinfo
  {journal} {Applied Physics Letters},\ }\textbf {\bibinfo {volume} {94}},\
  \bibinfo {pages} {162106} (\bibinfo {year} {2009})},\ \Eprint
  {http://arxiv.org/abs/http://dx.doi.org/10.1063/1.3123391}
  {http://dx.doi.org/10.1063/1.3123391} \BibitemShut {NoStop}%
\end{thebibliography}%



\end{document}